\documentclass[aps,prc,reprint,groupedaddress]{revtex4-1}
\usepackage{ulem}
\usepackage{amsmath}
\usepackage{graphicx}
\usepackage{hyperref}
\usepackage{lineno}

\hypersetup{
    colorlinks = true,
    linkcolor = blue,
    citecolor = blue,
    urlcolor = blue,
    bookmarksnumbered = true,
    linktoc = all    
}
\begin{document}
\title{Spin Femtoscopy: A Framework for Revealing Genuine Spin Correlations}

\author{Kehao Zhang}
\author{Xuan Wang}
\author{Xiaofeng Luo}
\thanks{xfluo@ccnu.edu.cn} 
\affiliation{Key Laboratory of Quark \& Lepton Physics (MOE) and Institute of Particle Physics, \\ Central China Normal University}

\begin{abstract}

Spin correlations are among the most fundamental quantum observables in many-body systems, yet they remain difficult to access experimentally in relativistic heavy-ion collisions. Existing spin measurements, including hyperon polarization and vector-meson spin alignment, have revealed important single-particle spin phenomena, but genuine two-particle spin correlations in the produced hadronic system remain largely unexplored. Here we propose spin femtoscopy, a framework for accessing genuine two-particle spin correlations through spin-resolved femtoscopic measurements. The key principle is that different two-particle spin configurations can give rise to different femtoscopic correlation functions because of quantum statistics and spin-dependent final-state interactions. Using $\Lambda\Lambda$ pairs as a proof of principle, we exploit the self-analyzing weak decay of $\Lambda$ hyperons to construct spin-sensitive femtoscopic correlation functions with different singlet and triplet admixtures. We show that these observables provide experimental access to the spin-state populations of the pair and allow genuine spin correlations to be separated from spin-dependent femtoscopic mixing caused by quantum statistics and final-state interactions. This work extends femtoscopy from a probe of source geometry and final-state interactions to a framework for revealing the quantum spin structure of strongly interacting matter.

\end{abstract}

\maketitle

\section{Introduction}
In relativistic heavy-ion collisions, strongly interacting matter is created at extreme temperature and density, followed by rapid expansion, hadronization, and kinetic freeze-out. Over the past several decades, a broad set of correlation observables has been developed to study this evolution, including collective flow, event-by-event fluctuations, balance functions, jet correlations, and femtoscopic two-particle correlations. These measurements have revealed detailed information on the space-time structure and transport properties of the matter produced in high-energy nuclear collisions~\cite{PHENIX:2004vcz,STAR:2005gfr,ALICE:2022wpn,Arslandok:2023utm,Chen:2024aom}.

Quantum correlations provide one of the most direct manifestations of the microscopic structure of many-body systems~\cite{Amico:2007ag,Horodecki:2009zz,DeChiara:2018biz,Oliva:2026wbo}. Spin, however, remains a comparatively less explored quantum degree of freedom. The observation of global $\Lambda$ polarization and vector-meson spin alignment established that spin-dependent phenomena  provide new insight into vorticity, magnetic fields, hadronization, and the quantum structure of the produced medium~\cite{Liang:2004ph,STAR:2017ckg,Becattini:2016gvu,Huang:2020dtn,Fu:2020oxj,Chen:2024hki,STAR:2022fan,Chen:2024afy}. These developments have opened a rapidly growing field of spin physics in heavy-ion collisions. Nevertheless, most existing measurements focus on single-particle spin observables or collective spin alignment. More recently, attention has turned to two-particle systems, such as $\Lambda\Lambda$ and $\Lambda\bar{\Lambda}$ pairs, where spin correlations can provide direct information on the joint spin state of the pair~\cite{Pang:2016igs,Sheng:2022wsy,Lv:2024uev,STAR:2025njp}. However, genuine two-particle spin correlations remain experimentally challenging. The difficulty is fundamental. In most hadronic final states, particle spin is not measured directly on an event-by-event basis. Momentum, charge, baryon number, and strangeness can be reconstructed with high precision, but the spin state of a particle pair is usually hidden. Moreover, the widely used experimental observables for two-particle spin correlations follow a strategy similar to femtoscopic measurements, but typically focus on the opening-angle distributions of daughter particles from weak decays. Such observables, however, can be affected by a variety of two-particle final-state effects, including quantum statistics, spin-dependent strong interactions, coupled-channel dynamics, resonance contributions, and other correlations associated with particle production. Consequently, signals of genuine spin correlations may be obscured or contaminated by these effects, particularly in small collision systems where hadronic final-state interactions and coupled-channel contributions can be sizable. We therefore refer to spin correlations originating from the earlier stages of the collision, for example, during partonic evolution or quark hadronization, rather than from two-particle final-state effects, as genuine spin correlations. A generic framework that can disentangle these contributions and provide direct experimental access to genuine two-particle spin correlations is thus essential.

In this work, we introduce the concept of spin femtoscopy, which enables a novel experimental approach to measure two-particle spin correlations through conventional femtoscopic techniques. By exploiting the self-analyzing weak decay of $\Lambda$ hyperons, spin information is incorporated into femtoscopic correlation measurements, allowing the construction of spin-resolved correlation functions corresponding to different spin-state compositions. The key challenge addressed in this work is that an observed angular correlation of weak-decay products is not automatically equivalent to a genuine spin correlation of the parent baryons. In an identical-fermion system, different spin states may have different femtoscopic responses because of quantum statistics and spin-dependent final-state interactions. As a result, the measured decay-angle distribution can acquire a nontrivial angular dependence. Spin femtoscopy provides a framework to disentangle these effects by treating the spin-state population and the femtoscopic correlation function simultaneously. This framework can serve as a new tool to separate genuine two-particle spin correlations from spin-dependent femtoscopic mixing in relativistic heavy-ion collisions.

The paper is organized as follows. In Sec. II, we introduce the basic spin-femtoscopic framework and discuss how different two-particle spin configurations can lead to different femtoscopic correlation functions. We then show how the self-analyzing weak decay of $\Lambda$ hyperons enables spin-sensitive measurements in the $\Lambda\Lambda$ system, and how spin-dependent femtoscopic responses can bias conventional extractions of spin correlations if not properly accounted for. In Sec. III, we validate the extraction methods using AMPT-based toy-model simulations. In Sec. IV, we discuss the implications of this framework for the study of the $\Lambda\Lambda$ interaction, the search for the $H$ dibaryon, and coupled-channel dynamics in the $S=-2$ sector. Finally, we summarize the main conclusions and outline future opportunities for applying spin femtoscopy to the study of genuine quantum spin correlations in heavy-ion collisions.

\section{Spin-femtoscopic observables and physical interpretation}

\subsection{Correlation function and the \texorpdfstring{$\Lambda\Lambda$}{LambdaLambda} interaction}

Two-particle femtoscopy probes the space-time structure of particle emission and the low-energy interaction between hadrons through the momentum correlation function. Experimentally, the correlation function is constructed as the ratio of the pair distribution from the same event to that from mixed events. In the Koonin--Pratt formalism, the correlation function is written as

\begin{equation}
C(q)=\int d^{3}r\,S(\mathbf{r})\,\left|\psi_{\mathbf{q}}(\mathbf{r})\right|^{2},
\end{equation}

where $\mathbf{q}$ is the relative momentum of the particle pair in the pair rest frame, and $\mathbf{r}$ denotes their relative separation at the time of emission. Here, $S(\mathbf{r})$ is the relative source function, while $\psi_{\mathbf{q}}(\mathbf{r})$ is the relative two-particle wave function~\cite{Pratt:1986cc,Lisa:2005dd}. The latter contains both the effects of quantum statistics (QS) and the strong final-state interaction (FSI), making the correlation function sensitive to the low-energy scattering properties of the particle pair.

For the $\Lambda\Lambda$ system, the two particles are identical spin-$1/2$ fermions. The total two-particle wave function must therefore be antisymmetric under particle exchange,
\begin{equation}
\Psi_{\rm total}(1,2)=-\Psi_{\rm total}(2,1).
\end{equation}
The spin-singlet ($J=0$) wave function is antisymmetric under exchange and must be accompanied by a symmetric spatial wave function, whereas the spin-triplet ($J=1$) wave functions are symmetric and require antisymmetric spatial wave functions. Thus, the spin and spatial exchange symmetries are constrained by the Pauli principle. For an ensemble with uncorrelated spins at production, the singlet and triplet channels have statistical weights of $1/4$ and $3/4$, respectively.

\begin{figure}[t]
\centering
\includegraphics[width=\columnwidth]{ 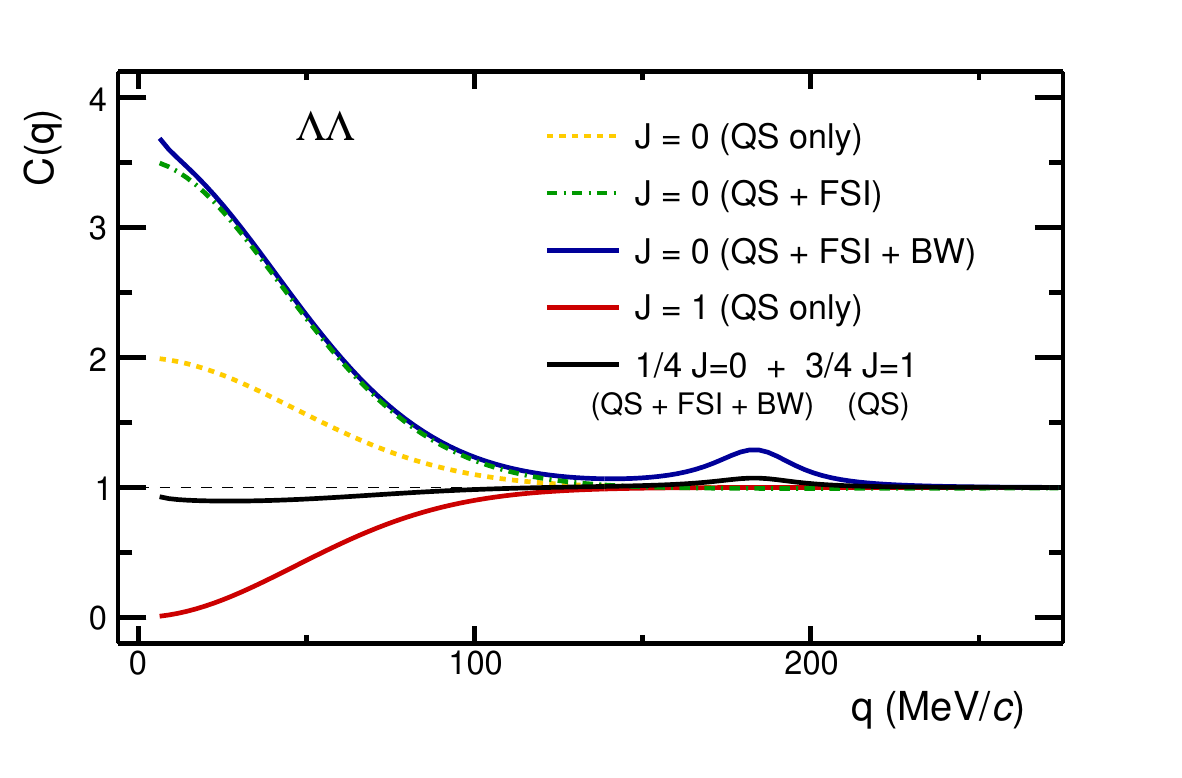}
\caption{
Theoretical $\Lambda\Lambda$ correlation functions for different spin configurations. The yellow, green, and blue curves correspond to the $J=0$ channel including quantum statistics (QS) only, QS together with the $\Lambda\Lambda$ final-state interaction (FSI), and QS+FSI with an additional Breit--Wigner (BW) resonance contribution, respectively. The red solid curve shows the $J=1$ channel including only the quantum-statistical effect. The black solid curve represents the spin-averaged correlation function,
$\frac{1}{4}C_{J=0}+\frac{3}{4}C_{J=1}$,
corresponding to an unpolarized $\Lambda\Lambda$ sample. The correlation functions are calculated within the Lednick\'y--Lyuboshits formalism assuming a Gaussian source with radius $r_{0}=1.5$~fm. The $\Lambda\Lambda$ interaction is described using the HAL QCD scattering parameters $f_{0}=0.99$~fm and $d_{0}=4.9$~fm. The BW contribution corresponds to a hypothetical near-threshold resonance located 30~MeV above the $\Lambda\Lambda$ threshold with a width of 15~MeV.
}
\label{fig:theory_cf}
\end{figure}
Near the $\Lambda\Lambda$ threshold, the relative motion is dominated by the $S$ wave ($L=0$). Since the $S$-wave spatial wave function is symmetric, only the spin-singlet ($J=0$) channel is allowed by the Pauli principle. As a consequence, the dominant strong $\Lambda\Lambda$ interaction is expected to arise from the $J=0$ channel. Contributions from $J=1$ channel are suppressed and neglected. In the present calculations, the $J=1$ channel is affected solely by quantum statistics. Any near-threshold resonance or bound state, such as the hypothesized $H$ dibaryon~\cite{Jaffe:1976yi}, is also expected to manifest itself only in the $J=0$ correlation function.

Figure~\ref{fig:theory_cf} illustrates the expected correlation functions calculated within the Lednick\'y--Lyuboshits (LL) framework~\cite{Lednicky:1981su}. In this approach, the two-particle relative wave function is constructed as a superposition of a free outgoing wave and a scattered wave, while the source function is assumed to follow a Gaussian form with a characteristic radius $r_0 = 1.5~\mathrm{fm}$. The final-state interaction is implemented through the low-energy effective-range expansion of the scattering amplitude, where the leading parameters are the scattering length $f_{0}$ and the effective range $d_{0}$. In this work, we adopt the HAL QCD-inspired values $f_{0}=0.99~\mathrm{fm}$ and $d_{0}=4.9~\mathrm{fm}$~\cite{Inoue:2010es,HALQCD:2019wsz,Kamiya:2021hdb}. In the femtoscopy convention, a positive scattering length corresponds to an attractive interaction, while a negative scattering length indicates a repulsive interaction or the presence of a bound state. The adopted value $f_{0}=0.99~\mathrm{fm}$ therefore indicates a weakly attractive $\Lambda\Lambda$ interaction. In addition, a possible $H$-dibaryon contribution is modeled as a Breit--Wigner resonance located $30~\mathrm{MeV}$ above the $\Lambda\Lambda$ threshold with a width of $15~\mathrm{MeV}$\cite{Morita:2014kza}. The Breit–Wigner component is introduced only as an illustrative example of a spin-selective near-threshold structure, not as a fit to data or a unique prediction for the H-dibaryon.

The figure compares the following contributions: (i) the pure QS correlation in the $J=0$ channel; (ii) the $J=0$ correlation including both QS and FSI; (iii) the $J=0$ correlation including QS, FSI, and the Breit--Wigner resonance; (iv) the pure QS correlation in the $J=1$ channel; and (v) the spin-averaged correlation function,

\begin{equation}
C_{\rm ave}(q)
=
\frac{1}{4}C_{J=0}(q)
+
\frac{3}{4}C_{J=1}(q).
\end{equation}

The comparison clearly demonstrates that quantum statistics alone produces markedly different correlation functions for the singlet and triplet channels due to the different exchange symmetry of their spatial wave functions. The strong interaction further modifies only the $J=0$ channel, while the resonance contribution generates an additional peak in the correlation function.

\subsection{Spin-resolved correlation function}

The experimentally measured $\Lambda\Lambda$ correlation function is, in essence, a statistical average over the spin-singlet and spin-triplet channels. As a result, the characteristic features of the spin-singlet ($J=0$) contribution are significantly diluted by the dominant triplet component in the observed signal. Owing to the inability of conventional femtoscopy to distinguish different spin states, the strong-interaction parameters extracted from experimental correlation measurements likewise correspond to spin-averaged quantities. This spin averaging limits the ability of conventional femtoscopy to discriminate between different spin channels. Motivated by this limitation, it is necessary to introduce spin-sensitive observables that can disentangle different spin configurations and enhance the contribution from each spin channel.

\begin{figure}[t]
\centering
\includegraphics[width=\columnwidth]{ 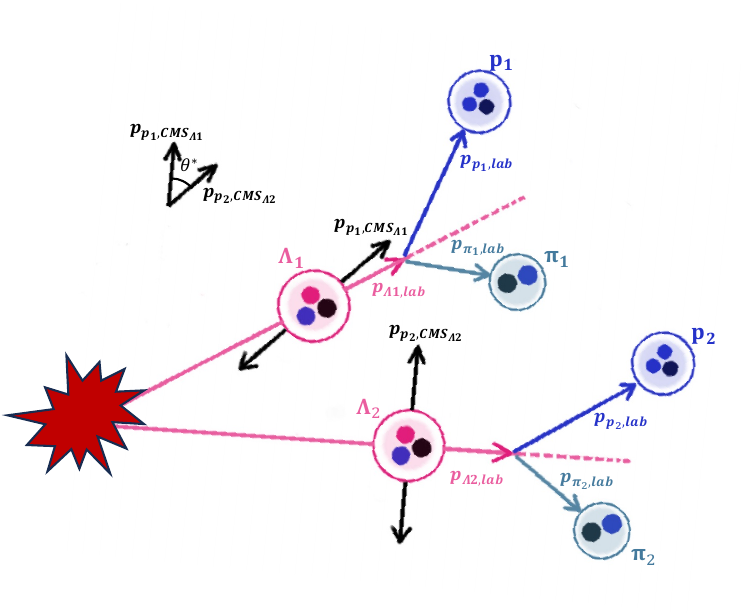}
\caption{
Schematic illustration of the determination of the opening angle $\theta^{*}$
between the decay protons in the respective rest frames of their parent
$\Lambda$ hyperons. The angular distribution of $\theta^{*}$ provides a
probe of the spin  correlation of the $\Lambda\Lambda$ pair.
}
\label{fig:theta_star}
\end{figure}

\begin{figure*}[htbp]
\centering
\includegraphics[width=0.85\textwidth]{ 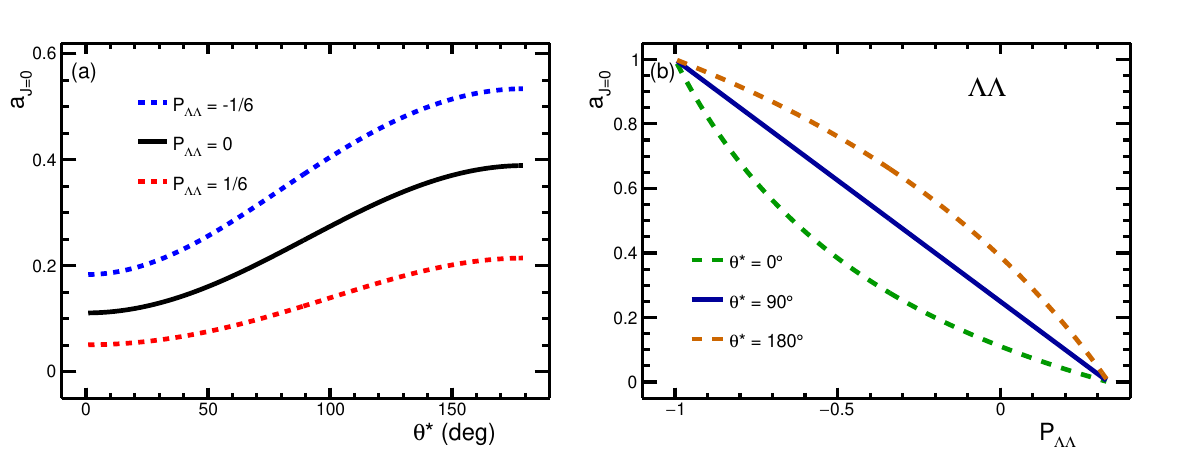}
\caption{
(a) Predicted spin-singlet fraction, $a_{J=0}$, as a function of the proton opening angle $\theta^{*}$ for different spin correlations. The black solid curve corresponds to the unpolarized case ($P_{\Lambda\Lambda}^{\rm gen}=0$), while the blue and red dashed curves represent $P_{\Lambda\Lambda}^{\rm gen}=-1/6$ and $P_{\Lambda\Lambda}^{\rm gen}=1/6$, respectively. Larger proton opening angles preferentially select the spin-singlet ($J=0$) channel, and the angular dependence of $a_{J=0}$ becomes more pronounced as the spin correlation deviates from zero. (b) Predicted spin-singlet fraction, $a_{J=0}$, as a function of the genuine spin correlation $P_{\Lambda\Lambda}^{\rm gen}$ for different proton opening angles $\theta^{*}$. The dark blue solid curve corresponds to $\theta^{*}=90^\circ$, while the dark green and dark yellow dashed curves represent $\theta^{*}=0^\circ$ and $\theta^{*}=180^\circ$, respectively.
}
\label{fig:spin_fraction}
\end{figure*}
Fortunately, the weak decay of the $\Lambda$ hyperon violates parity, allowing the polarization of the parent $\Lambda$ to be analyzed through the angular distribution of the decay proton~\cite{LeeYang1957},

\begin{equation}
\frac{dN}{d\cos\theta} \propto 1+\alpha_{\Lambda}P_{\Lambda}\cos\theta, \end{equation} 

where $\theta$ is the angle between the proton momentum in the $\Lambda$ rest frame and the $\Lambda$ polarization direction~\cite{STAR:2018gyt}, and $\alpha_{\Lambda}=0.746$ is taken from the 2026 Particle Data Group (PDG) recommended value for the weak decay asymmetry parameter of the $\Lambda$ hyperon~\cite{PDG2026}. Therefore, the proton momentum acts as a spin analyzer of the parent $\Lambda$. For a $\Lambda\Lambda$ pair, the decay protons are boosted into the respective rest frames of their parent $\Lambda$ hyperons, and the opening angle $\theta^{*}$ between the two boosted protons is determined, as illustrated in Fig.~\ref{fig:theta_star}. This relative angle $\theta^{*}$ carries information on the spin  correlation of the parent hyperons.

The spin correlation of a $\Lambda\Lambda$ pair can be characterized by~\cite{LednickyLyuboshitz2001}
\begin{equation}
P_{\Lambda\Lambda}
=
\frac{1}{3}
\left\langle
\hat{\boldsymbol{\sigma}}_1\cdot
\hat{\boldsymbol{\sigma}}_2
\right\rangle,
\end{equation}
where $\hat{\boldsymbol{\sigma}}_i$ is the Pauli spin operator acting on the spin space of the $i$-th $\Lambda$ hyperon, and the expectation value is evaluated over the two-particle spin ensemble. Since $\hat{\boldsymbol{\sigma}}_1\cdot\hat{\boldsymbol{\sigma}}_2$ has eigenvalues $-3$ and $+1$ for the spin-singlet and spin-triplet states, respectively, the spin correlation satisfies
\begin{equation}
-1\leq P_{\Lambda\Lambda}\leq\frac{1}{3}.
\end{equation}
The lower and upper bounds correspond to pure singlet and triplet states, respectively. For an uncorrelated spin ensemble with singlet and triplet statistical weights of $1/4$ and $3/4$, $P_{\Lambda\Lambda}=0$. Negative values indicate singlet enhancement, while positive values indicate triplet enhancement.

In the present framework, we distinguish between the genuine spin correlation $P_{\Lambda\Lambda}^{\rm gen}$ and the observed spin correlation $P_{\Lambda\Lambda}^{\rm obs}(q)$. The former refers to the spin correlation established at particle production, before applying spin-dependent femtoscopic weights, whereas the latter represents the effective spin correlation inferred from the decay angular distribution at a given relative momentum $q$ after quantum-statistical and final-state interaction effects are included.

In the absence of spin-dependent femtoscopic weighting, the angular distribution of the decay protons can be expressed as~\cite{Gong:2021bcp,Tornqvist1981,Tornqvist1986}
\begin{equation}
\frac{dN}{d\cos\theta^{*}}
\propto
1+\alpha_{\Lambda}^{2}
P_{\Lambda\Lambda}^{\rm gen}
\cos\theta^{*},
\label{tradition}
\end{equation}
where $\theta^{*}$ is the opening angle between the decay protons in the respective $\Lambda$ rest frames. In this limit, the angular coefficient reflects the genuine spin correlation at production. When spin-dependent femtoscopic effects are included, however, the observed angular distribution can be modified, and the extracted correlation need not equal $P_{\Lambda\Lambda}^{\rm gen}$.

For pure total-spin eigenstates, the angular distributions reduce to
\begin{equation}
\left.
\frac{dN}{d\cos\theta^{*}}
\right|_{J=0}
\propto
1-\alpha_{\Lambda}^{2}\cos\theta^{*},
\end{equation}
and
\begin{equation}
\left.
\frac{dN}{d\cos\theta^{*}}
\right|_{J=1}
\propto
1+\frac{1}{3}\alpha_{\Lambda}^{2}\cos\theta^{*}.
\end{equation}

The two spin channels therefore exhibit different angular dependences. The spin-singlet contribution is enhanced at large proton opening angles ($\theta^{*}\approx180^{\circ}$), whereas the spin-triplet contribution is favored at small opening angles ($\theta^{*}\approx0^{\circ}$).

For a $\Lambda\Lambda$ sample with vanishing spin correlation, the singlet and triplet states contribute with their statistical weights,

\begin{equation}
\frac{dN}{d\cos\theta^{*}}\propto\frac{1}{4}
\left(
1-\alpha_{\Lambda}^{2}\cos\theta^{*}
\right)
+
\frac{3}{4}
\left(
1+
\frac{1}{3}
\alpha_{\Lambda}^{2}\cos\theta^{*}
\right)
=1,
\end{equation}

resulting in a flat angular distribution. In this case, the fraction of spin-singlet pairs at a given opening angle is

\begin{equation}
a_{J=0}(\theta^{*})
=
\frac{
\frac14
\left(
1-\alpha_{\Lambda}^{2}\cos\theta^{*}
\right)
}
{
\frac14
\left(
1-\alpha_{\Lambda}^{2}\cos\theta^{*}
\right)
+
\frac34
\left(
1+\frac13\alpha_{\Lambda}^{2}\cos\theta^{*}
\right)
},
\end{equation}

which increases monotonically with $\theta^{*}$. Therefore, selecting different proton opening-angle intervals naturally changes the relative fractions of the $J=0$ and $J=1$ components.

More generally, let $w_0$ denote the spin-singlet fraction at production, with the corresponding triplet fraction $1-w_0$. Before including spin-dependent femtoscopic weighting, the angular distribution becomes
\begin{equation}
\begin{aligned}
\frac{dN}{d\cos\theta^{*}}\propto\;&
w_0\left(1-\alpha_{\Lambda}^{2}\cos\theta^{*}\right)\\
&+(1-w_0)\left(1+\frac13\alpha_{\Lambda}^{2}\cos\theta^{*}\right).
\end{aligned}
\end{equation}
Equivalently,
\begin{equation}
\frac{dN}{d\cos\theta^{*}}\propto
1+\alpha_{\Lambda}^{2}\left(\frac{1-4w_0}{3}\right)\cos\theta^{*}.
\end{equation}
Comparison with Eq.~(\ref{tradition}) gives
\begin{equation}
P_{\Lambda\Lambda}^{\rm gen}=\frac{1-4w_0}{3}.
\label{plala}
\end{equation}
This relation connects the genuine spin correlation with the initial spin-singlet fraction, rather than with a femtoscopically reweighted fraction.

The spin-singlet fraction selected at a given proton opening angle, before femtoscopic weighting, is
\begin{equation}
\begin{split}
&a_{J=0}(\theta^{*},P_{\Lambda\Lambda}^{\rm gen})\\
&=\frac{w_0(1-\alpha_{\Lambda}^{2}\cos\theta^{*})}{w_0(1-\alpha_{\Lambda}^{2}\cos\theta^{*})+(1-w_0)(1+\frac13\alpha_{\Lambda}^{2}\cos\theta^{*})}\\
&=\frac{1}{1+\frac{(1+P_{\Lambda\Lambda}^{\rm gen})(3+\alpha_{\Lambda}^{2}\cos\theta^{*})}{(1-3P_{\Lambda\Lambda}^{\rm gen})(1-\alpha_{\Lambda}^{2}\cos\theta^{*})}}.
\end{split}
\end{equation}
Figure~\ref{fig:spin_fraction}(a) shows the angular dependence of $a_{J=0}$ for different values of $P_{\Lambda\Lambda}^{\rm gen}$. Larger opening angles preferentially select the spin-singlet channel. Figure~\ref{fig:spin_fraction}(b) illustrates the dependence of $a_{J=0}$ on $P_{\Lambda\Lambda}^{\rm gen}$ at fixed opening angles. At a given angle, the singlet fraction increases as $P_{\Lambda\Lambda}^{\rm gen}$ decreases.

\begin{figure*}[htbp]
\centering
\includegraphics[width=0.85\textwidth]{ 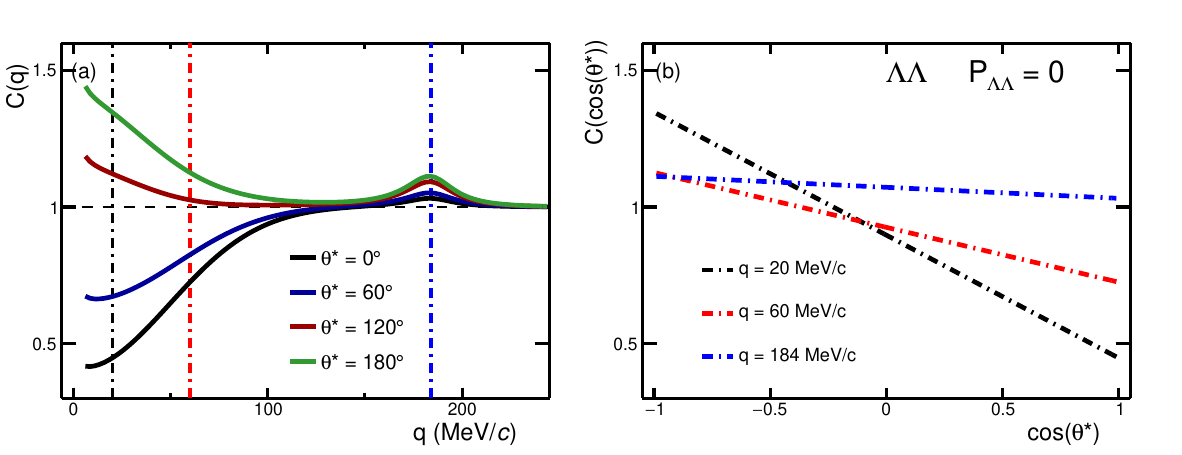}
\caption{
(a) $\Lambda\Lambda$ correlation functions calculated for $P_{\Lambda\Lambda}^{\rm gen}=0$ as a function of the relative momentum $q$ for different proton opening angles $\theta^{*}$. The black, blue, dark red, and green solid curves correspond to $\theta^{*}=0^\circ$, $60^\circ$, $120^\circ$, and $180^\circ$, respectively. The calculations are performed within the Lednick\'y--Lyuboshits formalism assuming a Gaussian source with radius $r_{0}=1.5$~fm. The $J=0$ channel includes quantum statistics (QS), the $\Lambda\Lambda$ final-state interaction (FSI), and a Breit--Wigner (BW) resonance contribution, while the $J=1$ channel includes only the QS effect. The $\Lambda\Lambda$ interaction is described using the HAL QCD scattering parameters $f_{0}=0.99$~fm and $d_{0}=4.9$~fm. The BW contribution corresponds to a hypothetical near-threshold resonance located 30~MeV above the $\Lambda\Lambda$ threshold with a width of 15~MeV. The vertical dashed lines indicate fixed relative momenta of $q=20$, $60$, and $184~\mathrm{MeV}/c$ (black, red, and blue, respectively). (b) The corresponding correlation functions as a function of $\cos\theta^{*}$ under the same model assumptions. The black, red, and blue curves correspond to $q=20$, $60$, and $184~\mathrm{MeV}/c$, respectively.
}
\label{fig:cf_theta}
\end{figure*}

In the present framework, we assume that the genuine spin correlation established at particle production is independent of the relative momentum $q$. Accordingly, $P_{\Lambda\Lambda}^{\rm gen}$, or equivalently the initial spin-singlet fraction $w_0$, is treated as a $q$-independent model parameter. This assumption concerns the spin composition at production and does not imply that the observed spin composition remains independent of $q$ after femtoscopic effects. Quantum statistics and final-state interactions modify the relative contributions of the two spin channels through their respective correlation functions, $C_{J=0}(q)$ and $C_{J=1}(q)$.

The correlation function integrated over the full proton opening-angle range is
\begin{equation}
C(q)=w_0C_{J=0}(q)+(1-w_0)C_{J=1}(q),
\label{eq:allrange}
\end{equation}
where $w_0$ represents the initial spin-singlet fraction, while the momentum dependence is contained in the spin-dependent correlation functions.

For a given proton opening angle, the correlation function becomes
\begin{equation}
\begin{split}
C(q,\theta^{*},P_{\Lambda\Lambda}^{\rm gen})=\;&
a_{J=0}(\theta^{*},P_{\Lambda\Lambda}^{\rm gen})C_{J=0}(q)\\
&+\left[1-a_{J=0}(\theta^{*},P_{\Lambda\Lambda}^{\rm gen})\right]C_{J=1}(q).
\end{split}
\end{equation}
For finite angular intervals, the coefficients are understood as the corresponding angularly averaged spin-state fractions.

Figure~\ref{fig:cf_theta}(a) presents the predicted correlation functions at different proton opening angles assuming $P_{\Lambda\Lambda}^{\rm gen}=0$. The $J=0$ channel includes quantum statistics, the $\Lambda\Lambda$ final-state interaction, and an illustrative Breit--Wigner resonance contribution, while the $J=1$ channel includes only the quantum-statistical effect. Although the genuine spin correlation vanishes, the correlation functions differ because the selected angular intervals contain different proportions of the two spin channels.

\subsection{Separating Genuine Spin Correlations from Femtoscopic Spin Mixing}

From an experimental perspective, the measurement of spin correlations and the construction of femtoscopic correlation functions are closely related, as both involve same-event and mixed-event pair distributions. In conventional spin-correlation measurements, mixed events are used to correct the decay angular distribution for detector acceptance~\cite{STAR:2014dcy,STAR:2025njp}. Equation~(\ref{tradition}) directly reflects the genuine spin correlation only when spin-dependent femtoscopic effects are absent.

When quantum statistics and final-state interactions are included, the double-differential angular distribution can be expressed as
\begin{equation}
\begin{aligned}
\frac{d^2N}{d\cos\theta^{*}dq}\propto\;&
w_0\left(1-\alpha_{\Lambda}^{2}\cos\theta^{*}\right)C_{J=0}(q)\\
&+(1-w_0)\left(1+\frac13\alpha_{\Lambda}^{2}\cos\theta^{*}\right)C_{J=1}(q),
\end{aligned}
\label{eq:spin_dependent_distribution}
\end{equation}
where $w_0$ is the initial spin-singlet fraction. The functions $C_{J=0}(q)$ and $C_{J=1}(q)$ describe the different femtoscopic responses of the two spin channels.

Equivalently, Eq.~(\ref{eq:spin_dependent_distribution}) can be written as
\begin{equation}
\frac{d^2N}{d\cos\theta^{*}dq}\propto
\left(1+\alpha_{\Lambda}^{2}P_{\Lambda\Lambda}^{\rm gen}\cos\theta^{*}\right)
C(q,\theta^{*},P_{\Lambda\Lambda}^{\rm gen}),
\label{eq:spin_dependent_factor}
\end{equation}
where $C(q,\theta^{*},P_{\Lambda\Lambda}^{\rm gen})$ is the angle-dependent femtoscopic correlation function defined above. As illustrated in Fig.~\ref{fig:cf_theta}(b), this function can exhibit a nonzero slope in $\cos\theta^{*}$ even for $P_{\Lambda\Lambda}^{\rm gen}=0$.

To distinguish the initial spin composition from the composition after femtoscopic weighting, we introduce the effective spin-singlet fraction at fixed $q$,
\begin{equation}
w_0^{\rm obs}(q)=\frac{w_0C_{J=0}(q)}{w_0C_{J=0}(q)+(1-w_0)C_{J=1}(q)}.
\label{eq:observed_singlet_fraction}
\end{equation}
This fraction describes the singlet contribution to the observed pair yield at fixed $q$ and generally differs from $w_0$ because the two spin channels have different femtoscopic responses. It is a reweighting of spin-channel yields and does not, by itself, imply microscopic spin rotation.

The double-differential distribution can therefore be rewritten as
\begin{equation}
\begin{aligned}
\frac{d^2N}{d\cos\theta^{*}dq}\propto\;&
\left[w_0C_{J=0}(q)+(1-w_0)C_{J=1}(q)\right]\\
&\times\left[1+\alpha_{\Lambda}^{2}P_{\Lambda\Lambda}^{\rm obs}(q)\cos\theta^{*}\right],
\end{aligned}
\label{eq:spin_correlation_distribution}
\end{equation}
where $P_{\Lambda\Lambda}^{\rm obs}(q)$ is the effective spin correlation inferred from the angular distribution at a given relative momentum. It satisfies
\begin{equation}
P_{\Lambda\Lambda}^{\rm obs}(q)=\frac{1-4w_0^{\rm obs}(q)}{3}.
\end{equation}
Using Eq.~(\ref{plala}), the observed correlation can be expressed in terms of the genuine correlation as
\begin{equation}
P_{\Lambda\Lambda}^{\rm obs}(q)=
\frac{1+P_{\Lambda\Lambda}^{\rm gen}-(1-3P_{\Lambda\Lambda}^{\rm gen})r(q)}{3(1+P_{\Lambda\Lambda}^{\rm gen})+(1-3P_{\Lambda\Lambda}^{\rm gen})r(q)},
\label{eq:observed_spin_correlation}
\end{equation}
where
\begin{equation}
r(q)=\frac{C_{J=0}(q)}{C_{J=1}(q)}.
\end{equation}
When $C_{J=0}(q)=C_{J=1}(q)$, the ratio $r(q)=1$ and $P_{\Lambda\Lambda}^{\rm obs}(q)=P_{\Lambda\Lambda}^{\rm gen}$. Conversely, different femtoscopic responses make the observed correlation momentum dependent even if the genuine correlation is independent of $q$. In particular, for $P_{\Lambda\Lambda}^{\rm gen}=0$,
\begin{equation}
P_{\Lambda\Lambda}^{\rm obs}(q)=\frac{1-r(q)}{3+r(q)},
\end{equation}
which can be nonzero. Thus, a nonzero observed spin correlation does not necessarily imply a nonzero genuine spin correlation at production. A coefficient fitted to a $q$-integrated angular distribution represents a pair-yield-weighted average and should not be confused with $P_{\Lambda\Lambda}^{\rm obs}(q)$ at a particular momentum.

To extract $P_{\Lambda\Lambda}^{\rm gen}$ in the presence of femtoscopic effects, we propose two complementary methods.

The first method exploits the angle-integrated correlation function. As shown in Eq.~(\ref{eq:allrange}), this function is a weighted sum of the spin-singlet and spin-triplet contributions, with $w_0$ as the initial spin-singlet fraction. By fitting the measured correlation function with sufficiently constrained spin-dependent correlation functions, $w_0$ can be extracted and $P_{\Lambda\Lambda}^{\rm gen}$ determined through Eq.~(\ref{plala}). Since this method does not require weak-decay angular information, it can in principle be extended to stable particles or particles without self-analyzing decay channels, provided the spin-dependent correlation functions can be adequately constrained.

The second method exploits the angular dependence of the spin-state fractions illustrated in Fig.~\ref{fig:spin_fraction}(a). Different values of $P_{\Lambda\Lambda}^{\rm gen}$ predict different angular dependences. By fitting femtoscopic correlation functions in different proton opening-angle intervals, the fraction of a selected spin state, such as $a_{J=0}(\theta^{*})$, can be extracted. Its angular dependence can then be compared with the theoretical expression to determine $P_{\Lambda\Lambda}^{\rm gen}$. This method explicitly uses the spin information encoded in the self-analyzing weak decays of the $\Lambda$ hyperons.

\section{Simulation based on AMPT Model}
\begin{figure*}[htbp]
    \centering
    \includegraphics[width=\textwidth]{ 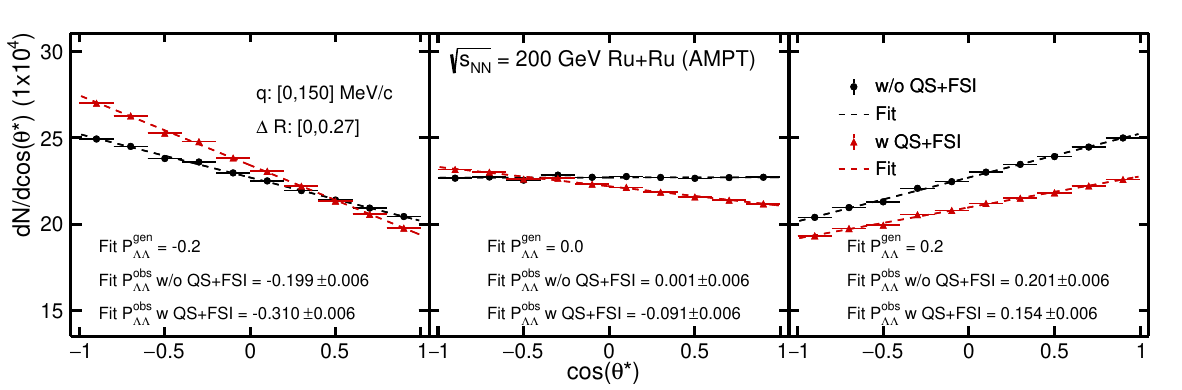}
    \caption{
    Angular distributions of decay protons for three input values of the genuine spin  correlation, $P_{\Lambda\Lambda}^{\rm gen}=-0.2$, $0$, and $0.2$ (from left to right), in $\sqrt{s_{NN}}=200$ GeV Ru+Ru collisions simulated with AMPT.
    The distributions are fitted using the conventional angular-distribution method, both without (blue) and with (red) quantum-statistical (QS) and final-state interaction (FSI) effects.
    The fitted angular-correlation coefficients are consistent with the input values when QS and FSI effects are absent, but deviate from them when these effects are included.
    }
    \label{fig:ampt_plalambda}
\end{figure*}
To validate the proposed methods for extracting the genuine $\Lambda\Lambda$ spin correlations in the presence of femtoscopic effects, we performed a simulation using events generated with the A Multi-Phase Transport (AMPT) model~\cite{Lin:2004en}. The simulation was configured for $\sqrt{s_{NN}}=200$ GeV Ru+Ru collisions, corresponding to the collision system and energy studied by the STAR experiment. AMPT was used to generate events and obtain the particle kinematic distributions. The spin correlation, quantum-statistical (QS) effects, and final-state interactions (FSI) were subsequently incorporated according to Eq.~\ref{eq:spin_dependent_distribution}. This procedure allowed us to construct samples with known input values of the genuine spin correlation, $P_{\Lambda\Lambda}^{\rm gen}$, and investigate whether the proposed methods could recover the underlying correlation.

\subsection{Conventional Angular-Distribution Method}

We first examined the conventional method based on the angular distribution of the decay protons. Three input values, $P_{\Lambda\Lambda}^{\rm gen}=-0.2$, $0$, and $0.2$, were considered. For each input, the angular distributions were analyzed both without and with QS and FSI effects.

The distributions were fitted with Eq.~(\ref{tradition}), and the extracted values are shown in Fig.~\ref{fig:ampt_plalambda}. Without QS and FSI, the fitted values reproduce the input values within statistical uncertainties. When QS and FSI are included, however, the fitted values deviate from the inputs. This deviation arises because the spin-dependent femtoscopic correlation modifies the observed angular distribution, such that its slope no longer reflects the genuine spin  correlation alone. This result demonstrates that the conventional angular-distribution method can be biased by femtoscopic effects and motivates the application of the two methods described below.

\subsection{Spin Femtoscopy Method 1: Angle-Integrated Correlation Function}

\begin{figure}[htbp]
    \centering
    \includegraphics[width=\columnwidth]{ 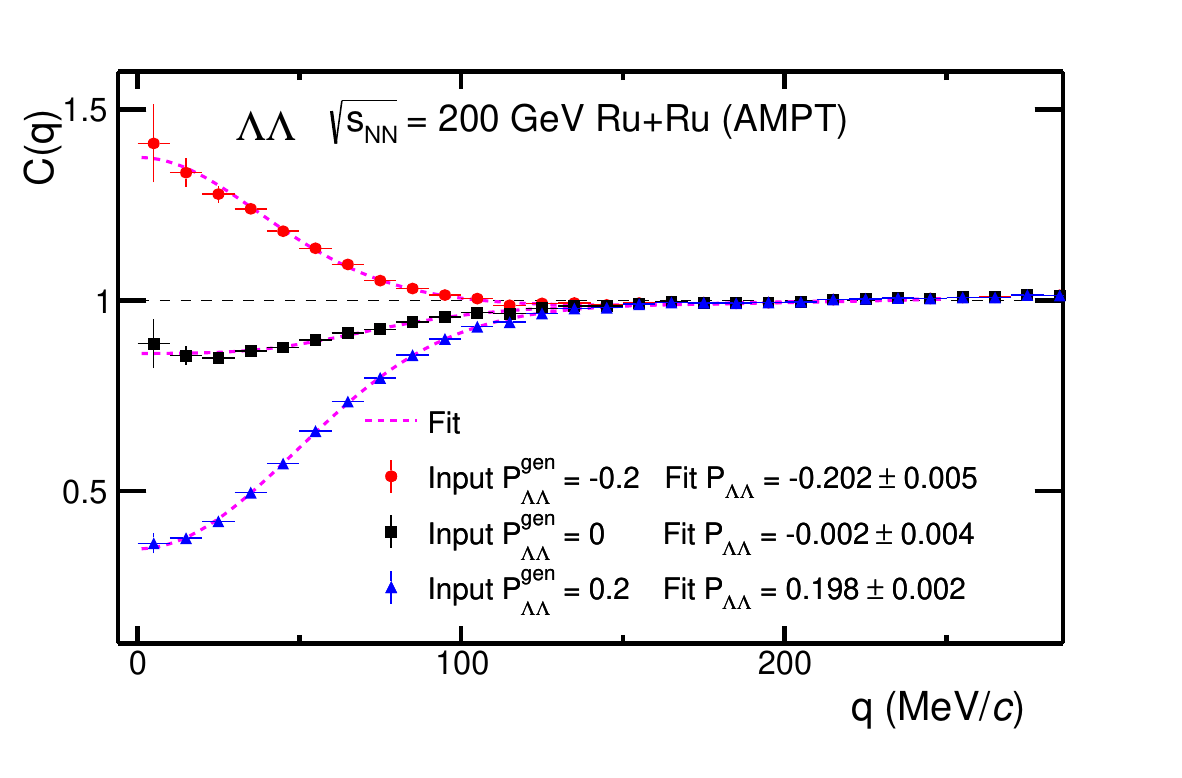}
    \caption{Angle-integrated $\Lambda\Lambda$ correlation functions as a function of relative momentum $q$ for three input values of the genuine spin  correlation, $P_{\Lambda\Lambda}^{\rm gen}=-0.2$, $0$, and $0.2$, in $\sqrt{s_{NN}}=200$ GeV Ru+Ru collisions simulated with AMPT. The symbols represent the simulated correlation functions, and the curves represent the corresponding fits using Eq.~(\ref{eq:allrange}). The extracted values of $P_{\Lambda\Lambda}^{\rm gen}$ are consistent with the input values within statistical uncertainties.}
    \label{fig:ampt_method1}
\end{figure}

For Method 1, the simulated correlation functions were integrated over the full proton opening-angle range and fitted using Eq.~(\ref{eq:allrange}) to extract the spin-singlet fraction $w_0$ for each input value of $P_{\Lambda\Lambda}^{\rm gen}$. The genuine spin  correlation was then determined using Eq.~(\ref{plala}). The resulting correlation functions and corresponding fits are shown in Fig.~\ref{fig:ampt_method1}.

The extracted values of $P_{\Lambda\Lambda}^{\rm gen}$ are $-0.202\pm0.005$, $-0.002\pm0.004$, and $0.198\pm0.002$ for the input values of $-0.2$, $0$, and $0.2$, respectively. All extracted values are consistent with the corresponding inputs within statistical uncertainties, demonstrating that Method 1 successfully recovers the genuine spin  correlation in this toy-model study.

\subsection{Spin Femtoscopy Method 2: Angle-Dependent Correlation Functions}
\begin{figure*}[t]
    \centering
    \includegraphics[width=\textwidth]{ 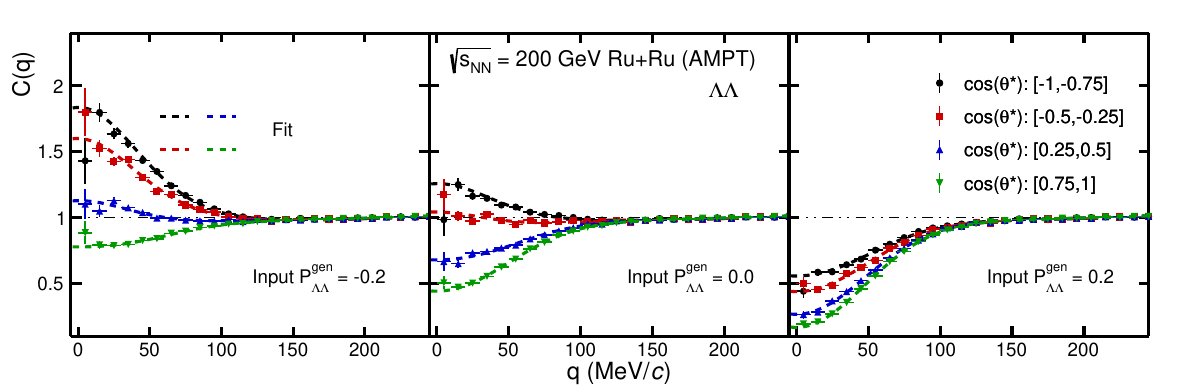}
    \caption{Angle-dependent $\Lambda\Lambda$ correlation functions for three input values of the genuine spin  correlation, $P_{\Lambda\Lambda}^{\rm gen}=-0.2$, $0$, and $0.2$ (from left to right), in $\sqrt{s_{NN}}=200$ GeV Ru+Ru collisions simulated with AMPT. Different symbols represent four selected $\cos\theta^{*}$ intervals, $[-1,-0.75]$, $[-0.5,-0.25]$, $[0.25,0.5]$, and $[0.75,1]$. The dashed curves represent the corresponding fits to extract the spin-singlet fractions. Only four representative intervals are shown for clarity, while all eight intervals are used in the analysis.}
    \label{fig:ampt_method2_cf}
\end{figure*}
For Method 2, the simulated correlation functions were fitted separately in different proton opening-angle intervals to extract the spin-singlet fraction $a_{J=0}(\theta^{*})$, using the spin-dependent correlation function introduced previously. Figure~\ref{fig:ampt_method2_cf} shows the correlation functions and corresponding fits for three input values of $P_{\Lambda\Lambda}^{\rm gen}=-0.2$, $0$, and $0.2$. For clarity, only four representative angular intervals are displayed, although all eight intervals were used in the analysis.

\begin{figure}[htbp]
    \centering
    \includegraphics[width=\columnwidth]{ 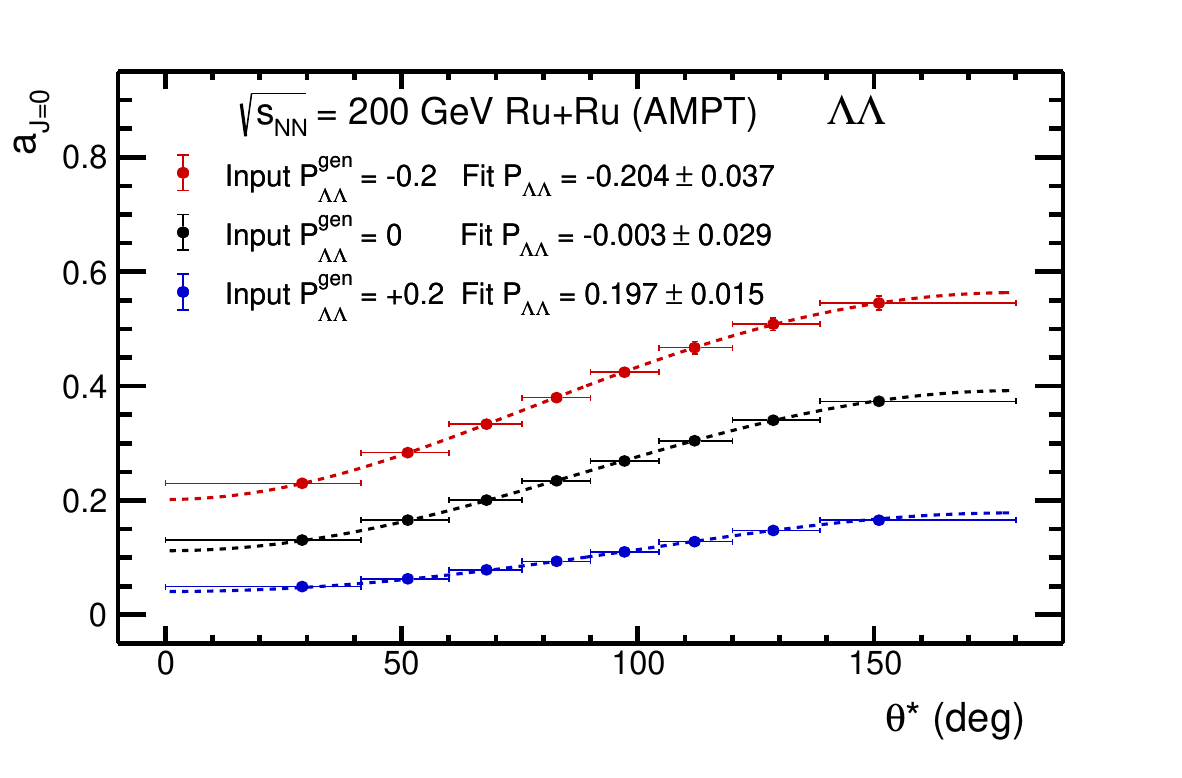}
    \caption{Extracted spin-singlet fraction $a_{J=0}$ as a function of the proton opening angle $\theta^{*}$ for three input values of $P_{\Lambda\Lambda}^{\rm gen}=-0.2$, $0$, and $0.2$ in $\sqrt{s_{NN}}=200$ GeV Ru+Ru collisions simulated with AMPT. The symbols represent the fractions extracted from the angle-dependent correlation functions, and the dashed curves represent fits using the theoretical angular dependence. The extracted values of $P_{\Lambda\Lambda}^{\rm gen}$ agree with the corresponding inputs within statistical uncertainties.}
    \label{fig:ampt_method2_fraction}
\end{figure}

The extracted spin-singlet fractions as a function of $\theta^{*}$ are shown in Fig.~\ref{fig:ampt_method2_fraction}. The angular dependence was fitted with the theoretical expression for $a_{J=0}(\theta^{*},P_{\Lambda\Lambda}^{\rm gen})$ to determine the genuine spin correlation. The extracted values are $-0.204\pm0.037$, $-0.003\pm0.029$, and $0.197\pm0.015$ for the input values of $-0.2$, $0$, and $0.2$, respectively. These results are consistent with the corresponding inputs within statistical uncertainties, demonstrating that Method 2 can successfully recover the genuine spin  correlation.

Only statistical uncertainties from the finite simulated samples are considered in this toy-model study. Both proposed methods recover the input genuine spin correlations, whereas the conventional angular-distribution method exhibits deviations when QS and FSI effects are included. These results establish closure under the assumptions of the toy model; experimental applications require further study of uncertainties in the source function, spin-dependent interactions, and possible coupled-channel contributions.

\section{Implications for $H$ Dibaryon Searches and Coupled-Channel Dynamics}

The existence of the $H$ dibaryon has remained an open question since it was first proposed by Jaffe as a deeply bound six-quark state with strangeness $S=-2$ and spin $J=0$. Over the past several decades, extensive experimental searches and theoretical studies have not yet established its existence, but have gradually shifted the focus from the original deeply bound scenario toward a weakly bound state below the $\Lambda\Lambda$ threshold or a near-threshold resonance above it, should the $H$ dibaryon exist~\cite{Bashinsky:1997qv,Takahashi:2001nm,KEK-PSE224:1998trj,Yoon:2007aq,Wu:2026ync}.

As demonstrated in the previous section, Fig.~\ref{fig:spin_fraction}(a) shows that the spin-singlet fraction $a_{J=0}$ exhibits a clear dependence on the proton opening angle $\theta^{*}$. Large opening angles preferentially select the $J=0$ configuration, while small opening angles are dominated by the triplet contribution. Furthermore, Fig.~\ref{fig:cf_theta}(a) indicates that the signal associated with the $H$ dibaryon becomes more pronounced as $\theta^{*}$ increases. Therefore, instead of measuring only the spin-averaged correlation function, one can compare correlation functions or invariant-mass spectra in different $\theta^{*}$ intervals to selectively enhance the $J=0$ component. Consequently, a resonance or bound-state signal associated with the $H$ dibaryon may become more pronounced in correlation functions enriched with the spin-singlet component, potentially improving the sensitivity of experimental searches. This spin-resolved analysis provides a direct experimental approach to improving the sensitivity of femtoscopic searches for the $H$ dibaryon.

The $\Lambda\Lambda$ interaction is strongly coupled to several nearby baryon--baryon channels, including $N\Xi$, $\Sigma\Sigma$, and $\Sigma^{*}\Sigma^{*}$. Recent coupled-channel calculations predict four poles in the $S=-2$, $I=0$, $J=0$ sector, extending from the $\Lambda\Lambda$ threshold to the $\Sigma^{*}\Sigma^{*}$ threshold~\cite{Hu:2026pyr}. These poles may generate distinct threshold structures in the $\Lambda\Lambda$ invariant-mass spectrum, including enhancements, cusps, or other line-shape modifications. Since the coupled-channel poles considered in these calculations belong to the $I=0$, $J=0$ sector, their contribution is directly correlated with the spin-singlet fraction selected by the proton opening angle. As the opening angle increases, the fraction of the $J=0$ component increases, leading to a progressively stronger coupled-channel contribution. Therefore, the proton opening angle provides a natural experimental handle for probing the underlying coupled-channel dynamics.

Interestingly, a recent presentation at the DIS2026 conference reported preliminary results from the CMS Collaboration~\cite{CMS}, showing a nonzero $\Lambda\Lambda$ spin correlation with a pronounced dependence on the pair angular separation $\Delta R$. As discussed above, the observed spin correlation $P_{\Lambda\Lambda}^{\rm obs}(q)$ depends on the relative momentum $q$ through the ratio $r(q)=C_{J=0}(q)/C_{J=1}(q)$, as described by Eq.~(\ref{eq:observed_spin_correlation}). This raises the possibility that the observed $\Delta R$ dependence may be influenced by the momentum dependence of spin-dependent femtoscopic effects.

The pair angular separation is defined as
\begin{equation}
\Delta R=\sqrt{(\Delta y)^2+(\Delta\phi)^2},
\end{equation}
where $\Delta y$ and $\Delta\phi$ denote the rapidity and azimuthal-angle differences between the two $\Lambda$ hyperons. For two $\Lambda$ hyperons of equal mass, the invariant mass $M_{\Lambda\Lambda}$ is related to the relative momentum $q$ in the pair rest frame by
\begin{equation}
M_{\Lambda\Lambda}^{2}=4(m_\Lambda^{2}+q^{2}).
\end{equation}
In terms of the single-particle transverse momenta and rapidity and azimuthal-angle differences, the same invariant mass can be written as
\begin{equation}
\begin{split}
4q^{2}=2\big[&
m_{T,1}m_{T,2}\cosh(\Delta y)\\
&-p_{T,1}p_{T,2}\cos(\Delta\phi)
-m_\Lambda^{2}\big],
\end{split}
\end{equation}
where $m_{T,i}=\sqrt{m_\Lambda^{2}+p_{T,i}^{2}}$. For small rapidity and azimuthal-angle differences, expansion to second order gives
\begin{equation}
\begin{split}
4q^{2}\simeq{}&
2(m_{T,1}m_{T,2}-p_{T,1}p_{T,2}-m_\Lambda^2)\\
&+m_{T,1}m_{T,2}(\Delta y)^2\\
&+p_{T,1}p_{T,2}(\Delta\phi)^2,
\end{split}
\end{equation}
The first term depends on the transverse-momentum difference and can remain nonzero even when $\Delta R=0$.

This relation demonstrates that the relative momentum $q$ is connected to the pair angular separation $\Delta R$ through the two-particle kinematics. Although $\Delta R$ does not uniquely determine $q$, different $\Delta R$ intervals can correspond to different relative-momentum distributions. Consequently, the momentum dependence of $P_{\Lambda\Lambda}^{\rm obs}(q)$, arising from differences between the spin-singlet and spin-triplet femtoscopic correlation functions, may contribute to the observed $\Delta R$ dependence of the $\Lambda\Lambda$ spin correlation.

Conventional low-energy $\Lambda\Lambda$ interactions are generally expected to have their strongest influence in the low-$q$ region, which may predominantly affect the observed spin correlation at small $\Delta R$. This effect was subsequently demonstrated at LHC energies in an AMPT simulation study~\cite{Fan:2026try}. In addition, coupled-channel interactions, such as $\Lambda\Lambda$--$N\Xi$ and $\Lambda\Lambda$--$\Sigma\Sigma$ couplings, can generate additional structures at higher relative momenta, extending the momentum range over which $C_{J=0}(q)$ and $C_{J=1}(q)$ differ. Through the resulting momentum dependence of $r(q)$, these coupled-channel effects may modify $P_{\Lambda\Lambda}^{\rm obs}(q)$ over a broader momentum range and potentially contribute to the spin-correlation trend observed over a wide range of $\Delta R$ by CMS. 

Therefore, the observed $\Delta R$ dependence may reflect not only genuine spin correlations established during particle production but also subsequent spin-dependent femtoscopic effects, including coupled-channel dynamics. Although $q$ and $\Delta R$ are not uniquely related, this interpretation provides a possible explanation for the broad $\Delta R$ dependence. A quantitative assessment requires dedicated calculations incorporating the relevant pair kinematics, experimental acceptance, and spin-dependent coupled-channel interactions.

\section{Summary and Outlook}

In this work, we have proposed a spin femtoscopy framework for revealing genuine quantum spin correlations in heavy-ion collisions. As a novel probe, spin femtoscopy extends conventional femtoscopy from the study of source geometry and final-state interactions to the investigation of quantum spin correlations. By exploiting the parity-violating decay $\Lambda \rightarrow p\pi^{-}$, the proton opening angle serves as an experimental handle for the underlying spin configuration of the $\Lambda\Lambda$ pair. We have shown that the spin-singlet ($J=0$) and spin-triplet ($J=1$) components exhibit distinct angular distributions, allowing their relative contributions to be continuously tuned by selecting different proton opening-angle intervals. This spin-resolved approach enables the construction of femtoscopic correlation functions with different spin compositions and provides a practical method for accessing genuine spin correlations beyond conventional spin-averaged femtoscopy.

In AMPT-based toy-model simulations with known input spin correlations, the conventional angular-distribution method yields biased results when spin-dependent QS and FSI effects are included. By contrast, both proposed femtoscopic methods recover the input genuine spin correlations within statistical uncertainties, establishing a proof-of-principle closure test under the model assumptions.

The proposed framework can be further applied to the study of the $\Lambda\Lambda$ interaction, including  searches for the $H$-dibaryon and the investigations of coupled-channel dynamics in the $S=-2$ sector. Since possible near-threshold dibaryon states and coupled-channel structures are expected to couple preferentially to specific spin channels, spin femtoscopy provides a natural way to enhance the relevant spin component and improve experimental sensitivity.

Looking ahead, this framework is fully compatible with existing femtoscopic techniques and can be readily implemented in experimental measurements. A more complete spin-femtoscopic program would become possible if future experiments could directly measure, or constrain, the spin orientations of final-state particles such as protons~\cite{Bai:2026syt}. Such capability would extend the present method beyond statistical spin analysis through weak decays and would allow spin-resolved femtoscopy to be applied to stable particles. With increasing experimental precision, it is expected to become a powerful tool for studying baryon-baryon interactions, searching for exotic dibaryons~\cite{Luo:2022mtp,Liu:2024uxn,Zhang:2025lfn}, and exploring quantum spin structure of strongly interacting matter.

\textit{Acknowledgements}
This work is supported in part by the National Natural Science Foundation of China under Grant No. 12525509, No. 12447102, and the National Key Research and Development Program of China under contract No. 2022YFA1604900, and the Fundamental Research Funds for the Central Universities (XJ2026000302).
\bibliography{reference.bib}
\end{document}